\documentclass[aps,onecolumn,superscriptaddress,showpacs,prl]{revtex4}
\usepackage{graphicx,amsfonts,amsmath,color,amsbsy,amssymb}
\usepackage{epstopdf}
\usepackage{appendix}
\begin{document}
\title{Equilibrium morphologies and force extension behavior for polymers with hydrophobic patches: Role of quenched disorder}
\author{Ankur Mishra}
\affiliation{Department of Metallurgical Engineering and Materials Science, Indian Institute of Technology Bombay, Powai, Mumbai - 400076, INDIA.}
\author{Ajay Singh Panwar}
\email{panwar@iitb.ac.in}
\affiliation{Department of Metallurgical Engineering and Materials Science, Indian Institute of Technology Bombay, Powai, Mumbai - 400076, INDIA.}
\author{Buddhapriya Chakrabarti}
\email{buddhapriya.chakrabarti@durham.ac.uk}
\affiliation{Department of Mathematical Sciences, Durham University, Durham, DH1 3LE, UK.}

\date{\today}
\begin{abstract}
Motivated by single molecule experiments on biopolymers we explore equilibrium morphologies and force-extension behavior of copolymers with hydrophobic segments using Langevin dynamics simulations. We find that the interplay between different length scales, namely, the persistence length $\ell_{p}$, and the disorder correlation length $p$, in addition to the fraction of hydrophobic patches $f$, play a major role in altering the equilibrium morphologies and mechanical response. In particular, we show a plethora of equilibrium morphologies for this system, \textit{e.g.} core-shell, looped (with hybridised hydrophilic-hydrophobic sections), and extended coils as a function of these parameters. A competition of bending energy and hybridisation energies between two types of beads determines the equilibrium morphology. Further, mechanical properties of such polymer architectures are crucially dependent on their native conformations, and in turn on the disorder realisation along the chain backbone. Thus, for flexible chains, a globule to extended coil transition is effected via a tensile force for all disorder realisations. However, the exact nature of the force-extension curves are different for the different disorder realisations. In contrast, we find that force-extension behavior of semi-flexible chains with different equilibrium configurations \textit{e.g.} core-shell, looped, \textit{etc.} reveal a cascade of force-induced conformational transitions.
\end{abstract}
\maketitle

\section{Introduction}
\label{sec:intro}

Biopolymers such as DNA, RNA, and proteins carry out a majority of cellular processes that are crucial to sustaining life~\cite{Alberts:02}. These processes are often associated with conformational changes of biomolecules. Several single molecule experiments, that probe mechanical properties of biomolecules, aim to relate primary structure to their native conformation and their function~\cite{Bustamante:11}.

Theoretical investigations on mechanical properties of proteins, DNA and other bio-macromolecules often use coarse-grained descriptions modeling structural components as non-linear elastic springs. Sequence heterogeneities are manifested as stiff and compliant elements along the chain backbone and are modeled by a site dependent bending modulus $\kappa_i$ (where $i$ corresponds to the lattice site). Statistical mechanical techniques rooted in the helix-coil model~\cite{Poland:70} and the worm-like chain model~\cite{Fisher:64,Marko:95} have been employed to explore stretching and bending response of biopolymers~\cite{Kierfeld:04,Chakrabarti:05,Chakrabarti:06}. These models however assume an annealed form of disorder, allowing for the inter-conversion between the ``helix'' and ``coil'' phases of the polymer chain as a function of either temperature or force. In the case of quenched disorder the following question arises. How does the mechanical response of a polymer chain with quenched disorder differ from that of an annealed one?

Motivated by this question we explore the mechanical properties of a polymer chain with hydrophobic and hydrophilic beads representing the quenched disorder using Langevin dynamics simulations. We investigate the stretching response of a polymer subject to equal and opposite tensile forces applied to its ends. In our simulations the polymer chain is composed of hydrophobic and hydrophilic segments with the ability to self-hybridize. Our main result is that the equilibrium configuration as well as the mechanical response of the polymer chain is determined both by the fraction of hydrophobic contacts, $f$, as well as their arrangement along the chain backbone. We consider three disorder configurations: (a) random, corresponding to hydrophobic segments uniformly distributed throughout the chain backbone, (b) periodic, where a repeat unit of hydrophobic beads are interspersed with hydrophilic ones at regular intervals, and (c) block copolymer, where the hydrophobic groups are clumped together at one end of the chain. A schematic representation of the disorder arrangements is shown in Fig.~\ref{fig:schematic}.

The problem is similar in spirit to investigations on~\textit{de novo} protein design of $A-B$ copolymers that give rise to folded configurations by Khokhlov and co-workers~\cite{Chertovich:00,Kriksin:02,Zheligovskaya:99,Khokhlov:99,vandanOever:99,Berezkin:03,Khokhlov:04,Chertovich:04}. The goal of their study was to identify arrangements of $A$ and $B$ molecules on the chain backbone that give rise to globular states akin to native protein conformations. In order to investigate this Khokhlov~\textit{et al.} equilibrated a bead-spring polymer in a poor solvent forming a globule. The atoms that form the core shell of this globular structure were  ``colored''~\cite{Zheligovskaya:99,Khokhlov:99} and the chain stretched out to its full length in order to note the position of the colored atoms along the chain backbone. A set of backbone atom arrangements that lead to optimal packing geometries was thus identified. A large body of literature dealing with equilibrium conformations of polymers with quenched disorder in context of proteins has been reviewed by Pande~\textit{et al.}~\cite{Pande:00}. Using simulations and theory, some other studies have explored both optimal disorder realizations for most compact packing~\cite{Lau:89,Yue:92} and collapse transitions in copolymers~\cite{Dasmahapatra:06,Dasmahapatra:07}.

In contrast, however, the mechanical response of chains with imposed quenched disorder to an applied tensile force has received limited attention~\cite{Shakhnovich:02,Shakhnovich2:02}. Techniques rooted in replica field theory have been employed to determine phase diagrams of stress induced unfolding of random polymer sequences. Globular, necklace like, and coil configurations have been reported~\cite{Shakhnovich:02,Shakhnovich2:02}. These studies however focus on flexible polymers and do not address equilibrium configurations or force extension behaviors of semi-flexible chains with a distribution of hydrophobic/hydrophilic beads.

Our study is novel from this standpoint,~\textit{i.e.} it explores the equilibrium morphologies and force vs.\ extension behaviors of flexible and semi-flexible polymers with different disorder realisations. Starting with a polymer chain of a given persistence length $\ell_{p}$, with a specified arrangement of hydrophobic segments, we equilibrate the chain and then apply a tensile force to unravel the structure.

The main results from this study can be summarized as follows: we observe a plethora of equilibrium shapes as a function of the hydrophobic fraction $f$, the periodicity length $p$, and persistence length $\ell_{p}$. In particular, we observe globular states that resemble a core shell morphology with the hydrophobic beads forming the core, a looped configuration with hybridised hydrophobic/hydrophilic segments, and an extended coil morphology. A phase map demarcating regions where such phases are stable is shown in Fig~\ref{fig:Phase-Map-100} and ~\ref{fig:Phase-Map-400}. We also observe a sharp first order transition in the chain size at a critical force $F_{c}$ for all three disorder realizations at a high value of the hydrophobic bead fraction $f$ (see Fig.~\ref{fig:FvsExt}). For a fixed hydrophobic fraction, $f$, (and $f > 0.5$) the critical force, $F_{c}$, for the block copolymer distribution is observed to be higher than that of the periodic and random distribution. Interestingly, the difference between the critical forces corresponding to periodic and random distributions is negligible. Further, the critical force is zero for all three distributions for low values of $f$. A non-zero critical force $F_c$ is observed above a threshold value of $f$ which is found to be the same for all three distributions. The nature of the transition is similar to a first order phase transition for the block copolymer distribution whereas it appears to be a crossover phenomena for the random and periodic ones. For semiflexible chains with looped internal structure, the force vs.\ extension behavior manifests itself in a series of conformational phase transitions.

Most biopolymers, including DNA and proteins are semi-flexible and often leads to complex shapes such as DNA toroids~\cite{vandenBroek:10}. AFM microscopy can be used to identify these shapes via the associated force induced conformational phase transitions. The simulations presented in this article provides an~\textit{in silico} tool to understand equilibrium shapes and associated conformational transitions of such naturally occurring polymers.

The remainder of the article is organized as follows. We provide the details of the simulation method in the following section. The Results and Discussion section is subdivided into two sub-sections, containing a discussion of equilibrium morphologies, and a description of the force vs.\ extension curves, respectively. Finally we provide perspectives of this work in the Summary section.

\section{Simulation Method}\label{sec:simulation}

\begin{figure}[ht]
\includegraphics[width=8cm]{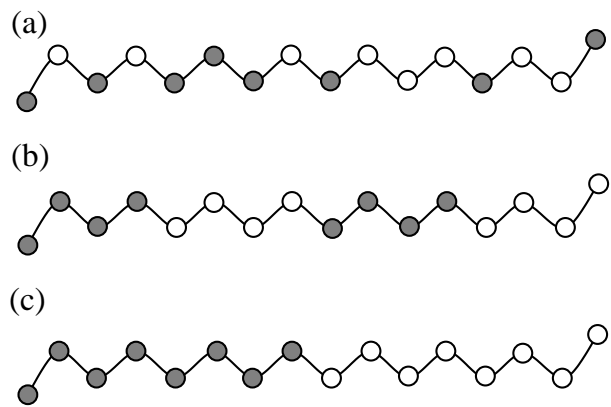}
\caption{Schematic representation showing hydrophobic groups (black circles) arranged along the chain backbone in (a) random, (b) periodic, and (c) block copolymer arrangements respectively. The fraction of hydrophobic beads is $f = 0.5$.}
\label{fig:schematic}
\end{figure}

The copolymer is modeled as a bead-spring chain consisting of $N$ beads linearly connected by $N-1$ springs. The copolymer comprises of two types of monomers (or beads), they are either hydrophobic or hydrophilic. As illustrated in Fig.~\ref{fig:schematic}, we consider three different arrangements of hydrophobic beads along the polymer chain. The hydrophobic beads are distributed either in a random, periodic or block copolymer arrangements. Figure~\ref{fig:schematic} represents the three different distributions for a chain with a fraction, $f = 0.5$, of hydrophobic groups.

Each bead in the polymer chain represents a Brownian particle, and its dynamics is described by the Langevin equation,

\begin{equation}
\label{eq:langevin}
m_{i} \frac{d^{2} {\bf r}_{i}}{d t^2} = - \zeta {\bf v}_{i} -\nabla_{i} U(r_{ij}) + {\bf F}^{r}_{i}(t) + {\bf F}^{ext}_{i},
\end{equation}

\noindent where $m_{i}$, ${\bf r}_{i}$ and ${\bf v}_{i}$ represent the mass, position and velocity of particle $i$, respectively. The random
force, ${\bf F}^{r}_{i}(t)$, arising from the bombardment of the monomer bead by the solvent molecules is described by the
fluctuation-dissipation theorem,

\begin{equation}
\label{eq:random}
\langle {\bf F}^{r}_{i}(t) {\bf F}^{r}_{j}(t^\prime) \rangle = 6 \zeta k_{B} T \delta(t - t^\prime) \delta_{ij},
\end{equation}

\noindent where $k_{B}$ is the Boltzmann constant, $T$ the absolute temperature, and $\zeta$ is the frictional drag. Externally applied, equal and opposite forces,  $F^{ext}_{i}$, act on the terminal beads of the chain. However, $F^{ext}_{i} = 0$ for the remaining polymer beads. Equation~\ref{eq:langevin} was solved numerically using the velocity-Verlet algorithm, with a time step $\Delta t = 0.001 \tau$, where $\tau = l_{0}(m / \epsilon_{LJ})^{1/2}$.

The net interaction potential, $U(r_{ij})$, is given by,

\begin{equation}
\label{eq:potential}
U(r_{ij}) = U_{LJ} + U_{bond} + U_{angle},
\end{equation}

\noindent corresponding to the excluded volume interactions, bond stretching, and the bending energy, respectively. The excluded volume interaction
between any two beads is described by,
\begin{eqnarray}
\label{eq:lj}
U_{LJ}(r_{ij}) = \left\{ \begin{array}{ll}
4\epsilon_{LJ}[{(\sigma/r_{ij})}^{12}-{(\sigma/r_{ij})}^{6}],
& r_{ij} \leq 2.5 \sigma, \\
0, \; r_{ij} > 2.5 \sigma,
\end{array} \right.
\end{eqnarray}

\noindent where $r_{ij} = | {\bf r}_{i} - {\bf r}_{j} |$, $\epsilon_{LJ}$ is the Lennard-Jones interaction parameter and $\sigma$ is the bead diameter. In our simulations, we have set the bead diameter $\sigma = 0.75 l_0$~\cite{Liu:02}, where $l_{0}$ is the bond length. We use $\epsilon_{LJ}$, $l_0$ and $m$ as scales for energy, length, and mass, respectively. The bond length, $l_{0}$, and the mass of a polymer bead, $m$ are set to a value of one. The bond stretching potential between adjacent beads is given by the FENE potential,
\begin{equation}
\label{eq:fene}
U_{bond}(r_{b}) = - \frac{1}{2} k r_{b}^2 \ln \left[1 - \left(r_{b}/R_{max}\right)^2 \right],
\end{equation}

\noindent where $r_{b}$ is the separation distance between adjacent beads, $R_{max}=1.5 l_{0}$ is the maximum allowable separation distance between bonded beads and $k$ is the spring constant. In order to model a semiflexible polymer, a bending potential is used and is given by,
\begin{equation}
\label{eq:wlc}
U_{angle}(\theta) = k_{\theta} \left[1 + cos \theta \right],
\end{equation}

\noindent where $k_{\theta}$ is the bending energy coefficient and $\theta$ is the angle between two adjacent bonds. We varied the values of $k_{\theta}$ between $0$ (a flexible chain with zero bending energy) and $30$ to vary the extent of semiflexibility of the polymer chain. Corresponding to a given value of $k_{\theta}$, a single polymer chain with all hydrophilic monomers was equilibrated, and a persistence length, $\ell_{p}$, was extracted from the decay of the tangent-tangent correlation function given by $\langle \hat{t}(0) \cdot \hat{t}(s) \rangle \sim \exp[- \vert s \vert/\ell_{p}]$~\cite{Marko:95}. Whereas, the lowest value of $k_{\theta} = 0$ corresponds to $l_{p} = 0.6$, an increase in $k_{\theta}$ results in an increase in $l_{p}$, and the highest used value of $k_{\theta} = 30$ corresponds to $l_{p} = 25$. For a value of $k_{\theta} = 0$, the polymer would be a freely-jointed chain, where the bond length would represent a Kuhn length, $b_{K}$, and satisfy $b_{K} = 2 l_{p}$. Then, for our model, $b_{K}=l_{0}=1$, and is nearly twice that of $l_{p} = 0.6$.

The simulations were carried out in the open-source molecular dynamics simulation package LAMMPS~\cite{Plimpton:95}. We used chain sizes of $N = 100, 200, 300$ and $400$ in our simulations. The polymer chain is enclosed inside a cubic box of edge length, $L_{box}$, (for instance, $L_{box} = 50$ $l_0$  for $N = 100$ and $L_{box} = 100$ $l_0$  for $N = 400$) with periodic boundary conditions imposed in all three directions. In order to model the appropriate effective solvent interactions, we set $\epsilon_{LJ} = 1.5$ for the hydrophobic beads and $\epsilon_{LJ} = 0.1$ for the hydrophilic ones. For a homopolymer, a choice of $\epsilon_{LJ} = 0.1$ for all beads simulates a polymer chain that demonstrates self-avoiding walk statistics, whereas a value of $\epsilon_{LJ} = 1.5$ results in a collapsed globule (results not shown here). When $\epsilon_{LJ} = 0.4$, we recover Gaussian statistics for the homopolymer. We varied the hydrophobic bead fraction,  $f$, between $0$ and $1$, to explore the effects of disorder on the force-extension response of the polymer chain.

The initial configuration of the chain was such that a fraction, $f$, of the beads were assigned as hydrophobic and the rest, $1-f$, were considered to be hydrophilic. Initially, the polymer chain is equilibrated in solution for a period of $4\times10^{6} - 4\times10^{7}$ time steps (depending on the chain size) in the absence of any external force ($F^{ext}=0$). The equilibration time chosen for a given chain length is such that it is several times larger than the corresponding Rouse relaxation time, thus ensuring that the chain has indeed equilibrated before the ensuing pulling cycle. The chain was then stretched from the equilibrium state to full extension by an incremental increase in pulling force, $F^{ext}$, from a value of $0.1$ to $10$. The pulling cycle is followed by a retraction cycle where $F^{ext}$ is decreased from $10$ to $0.1$. During the pulling and the retraction cycles, the polymer is equilibrated for a period of $5\times10^{6} - 2\times10^{7}$ time steps (again depending on the chain size) at a given value of $F^{ext}$. We calculate the time-averaged value of the root mean-square end-end distance, $\langle R^{2}\rangle^{1/2}$, over the equilibration period. Consequently, the extension of the polymer at that value of applied force is calculated. The data was averaged over $N_{run}=5$ disorder realizations for the three types of distributions and each value of $f$. In the case of the random distribution, the five realizations correspond to five statistically different arrangements of hydrophobic groups along the polymer backbone, for the same value of $f$. The results obtained from these different realizations were used to calculate the average values and standard deviations of the root mean-square end-end distance of the polymer corresponding to a particular value of force and fraction of hydrophobic groups

\section{\label{sec:results}Results and Discussion}

\subsection{\label{sec:eqmorph}Equilibrium morphologies}

\begin{figure}[ht]
\includegraphics[width=8cm]{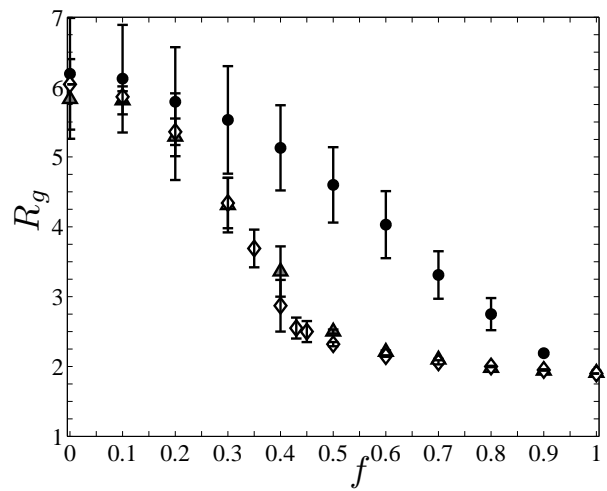}
\caption{Figure shows the average radius of gyration, $\langle R_g \rangle$, for a $N = 100$ chain over different disorder realizations as a function of the hydrophobic groups for (a) block copolymer (filled circles $\circ$), (b) periodic distribution (filled triangles $\triangle$), and random distribution (open diamonds $\lozenge$). The average is over $5$ different realizations for the random system.}
\label{fig:Rgvsf}
\end{figure}

Figure \ref{fig:Rgvsf} describes the variation of the radius of gyration, $R_{g}$, of a copolymer ($N=100$) as a function of $f$, for all three distributions at zero applied force. We find that as $f$ increases, $R_{g}$ decreases monotonically for all three cases. An increase in $f$ results in aggregation of hydrophobic groups leading to a monotonic decrease in $R_{g}$. The variation of $R_{g}$ with $f$ follows near identical trends for both the random and periodic distributions for the entire range of $f$. Both distributions display a sigmoidal behavior, where the polymer chains show a transition from a coiled state to a collapsed, globular state for $f > 0.4$. This is in contrast to the behavior of $R_{g}$ with $f$ for the block copolymer distribution. For $f \lesssim 0.2$, $R_{g}$ values for the block copolymer distribution are close to the ones corresponding to the periodic and random ones. However, for $0.2 \lesssim f \lesssim 0.8$, the rate of change of $R_{g}$ with respect to $f$ is lower than the other two
distributions. Also, the block copolymer does not go through a sharp phase transition as the other two distributions. For $f  \gtrsim 0.8$, $R_{g}$ values for all three distributions tend to a common value.  At $f = 1.0$, $R_{g}$ is the same for all three distributions.

In the case of a block copolymer, the hydrophobic groups at one end form a globule, while the rest of the chain assumes a coil structure. As $f$ increases, a greater fraction of the chain is in the globular state, whereas the hydrophilic block remains as a coil. Consequently, the entire chain does not form a compact globule for any value of $f$ apart from $f = 1$. However, for the periodic and random distributions, hydrophobic groups are distributed along the entire polymer chain, which enables attractive interactions between hydrophobic groups that are far separated along the chain contour. This results in a first order coil-to-globule transition at $f \approx 0.35$.

Thus, copolymers with $f > 0.35$ and zero bending energy organize themselves as compact globules in solution. However, the equilibrium structure can change if the bending energy cost (equivalently, the persistence length, $\ell_{p}$) increases. In addition, the equilibrium structure can also change if there is a change in the periodicity length, $p$, the length scale over which hydrophobic groups are repeated for a given value of $f$. Thus, the periodicity length, $p$, defines the disorder correlation length along the polymer contour. The choice of $p$ in our simulations is such that an integer number of periods make up the contour length, and a fraction $f$ of the length $p$ would comprise of hydrophobic beads, thus reflecting the overall hydrophobic fraction at that scale.

\begin{figure}[ht]
\includegraphics[width=10cm]{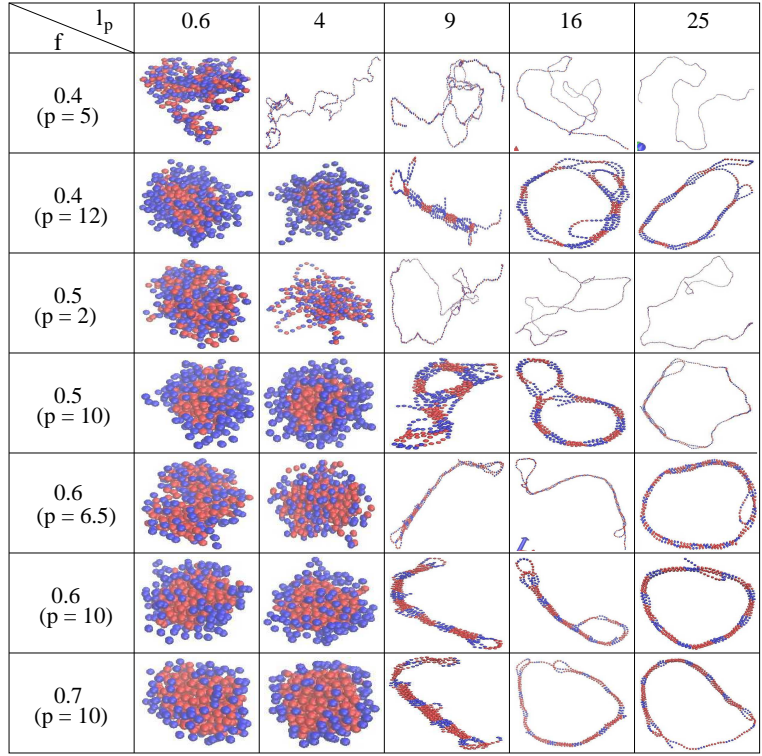}
\caption{Equilibrium morphologies of a $N = 400$ polymer chain as a function of different hydrophobic fractions, $f$, periodicity lengths, $p$, and persistence lengths, $\ell_{p}$. The hydrophobic and hydrophilic beads are indicated in red and blue colors, respectively. The chain rigidity/persistence length increases along each row while the hydrophobic fraction increases along each column. For flexible chains a core-shell like morphology is seen for all values of $f$, while for semi-flexible chains with large periodicity lengths, looped configurations are observed.}
\label{fig:Rep-Phase-400}
\end{figure}

Figure~\ref{fig:Rep-Phase-400} shows equilibrium morphologies of polymers (for $N = 400$) as a function of their persistence length $\ell_{p}$, hydrophobic fraction $f$, and periodicity length, $p$. For a semi-flexible chain, a combination of bending energy with the bending modulus $\kappa = k_B T \ell_{p}$, solvent quality $\epsilon_{LJ}$ and hybridisation energy between hydrophobic and hydrophilic segments dictates the equilibrium morphology. The table shows configuration snapshots with increasing bending modulus (or $\ell_{p}$) along each row and increasing hydrophobic fraction, $f$, as one moves down a column. It is seen that for small values of persistence length the equilibrium morphology resembles that of a core-shell like particle with the hydrophilic particles at the surface shielding an inner hydrophobic core. As the hydrophobic fraction increases, there are fewer hydrophilic particles to completely cover the globule surface, which exposes some of the hydrophobic beads to the solvent.

\begin{figure}[ht]
\includegraphics[width=10cm]{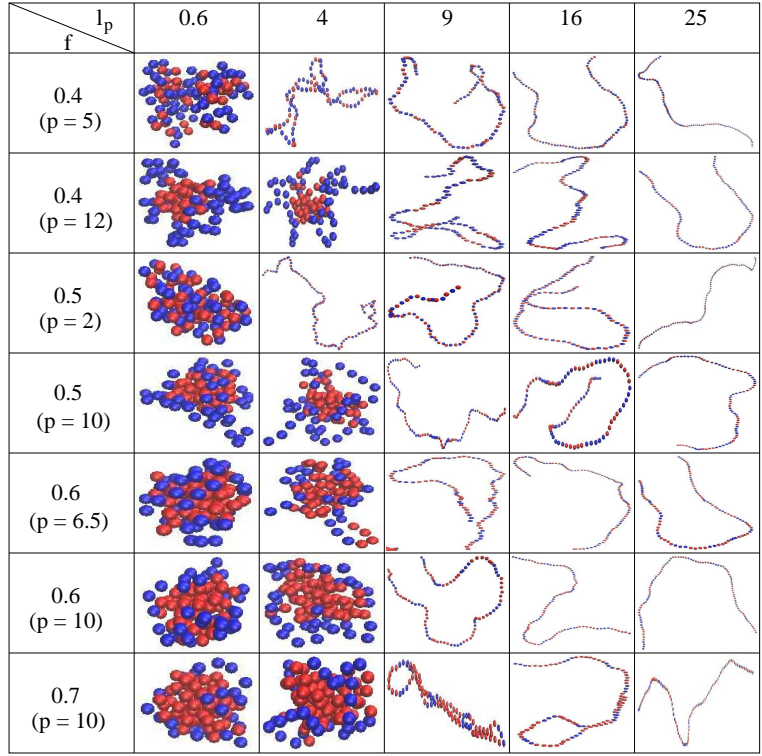}
\caption{Representative snapshots of dressed polymer architectures for $N = 100$ chain as a function of different hydrophobic fractions $f$, periodicity lengths $p$, and persistence length $\ell_{p}$. While core-shell morphologies similar to the $N = 400$ case is observed, looped configurations are rarely seen even for large persistence and periodicity lengths.}\label{fig:Rep-Phase-100}
\end{figure}

With increasing bending modulus it costs energy to bend the chain over length scales less than or comparable to the persistence length $\ell_{p}$. As a result the core-shell morphology is not seen. Instead the chain manifests itself in either an extended coil, an elongated fold or a loop like morphology. As is evident from Fig.~\ref{fig:Rep-Phase-400}, these newer structures appear over specific ranges of $f$ and $p$. For small values of both $f$ and $p$ ($f = 0.4, p = 5$ and $f = 0.5, p = 2$), a globular configuration for small $\ell_{p}$ gives way to a coil-like configuration with increasing $\ell_{p}$. For larger values of $f$ or $p$, elongated folds and loops are more common, and especially, dominate the lower right-hand side of Fig.~\ref{fig:Rep-Phase-400}. A polymer with a large $\ell_{p}$ value will bend over longer length scales, and if $p$ is also large, then its hydrophobic segments will hybridize over longer lengths along the contour and prefer to form an elongated fold or a loop. The occurence of these structures will also increase as $f$ increases, since hybridization of hydrophobic segments becomes much easier. Thus, the loop configuration is observed for large values of hydrophobic fractions, $f = 0.6, 0.7$, periodicities, $p = 6.5, 10$, and persistence lengths, $\ell_{p} = 16, 25$, respectively. It must be noted though that the ratio $\ell_{p}/L$ and not $\ell_{p}$ alone is the crucial parameter as it sets the scale over which sections of the chain can bend and hybridise. For example for $\ell_{p}/L \approxeq 10^{-3} - 0.0625$ sections of the chain which are a few persistence lengths apart can bend and hybridise. If however a smaller chain size is chosen \textit{e.g.} $N = 100$ as shown in Fig.~\ref{fig:Rep-Phase-100} such conformations are not energetically favorable and therefore not observed.

The representative configurations for a chain of contour length $N = 100$ and $\ell_{p}$ between $0.6 - 25$ (\textit{i.e.} $\ell_{p}/L \rightarrow 0.006 - 0.25$) are shown in Fig.~\ref{fig:Rep-Phase-100}. Two distinct configurations are seen here; core-shell structures similar to the $N = 400$ case for low values of $\ell_{p}$, and extended coils for large values of $\ell_{p}$. Although, simulations for both $N = 100$ and $N = 400$ were carried out for the same values of $\ell_{p}$, it is interesting to note that we do not observe any loop structures for $N = 100$, and fewer elongated folds for $N = 100$ when compared to $N = 400$. This observation further strengthens the argument that the ratio $\ell_{p}/L$ and not $\ell_{p}$ alone is the crucial parameter which determines the equilibrium structure of the copolymer.

\begin{figure}[ht]
\includegraphics[width=10cm]{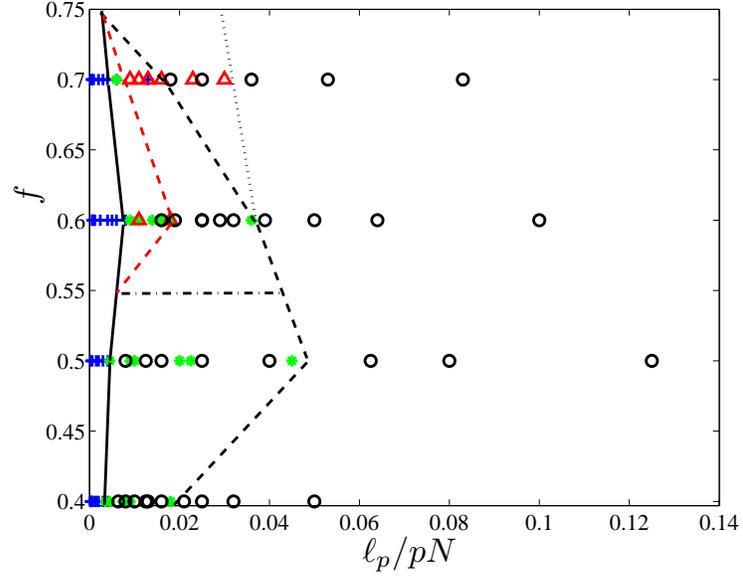}
\caption{Phase map for a $N = 100$ chain. The two key parameters that dictate the stability of different configurations is the hydrophobic fraction $f$, and the ratio of the persistence length to the periodicity length normalised by the total number of beads, $\ell_{p}/p N$. The different phases are indicated by globule ($+$), coil ($*$), folded ($\triangle$), and extended coils ($\circ$). The lines indicate $50 \%$ stability lines for the different phases.}\label{fig:Phase-Map-100}
\end{figure}

\begin{figure}[ht]
\includegraphics[width=10cm]{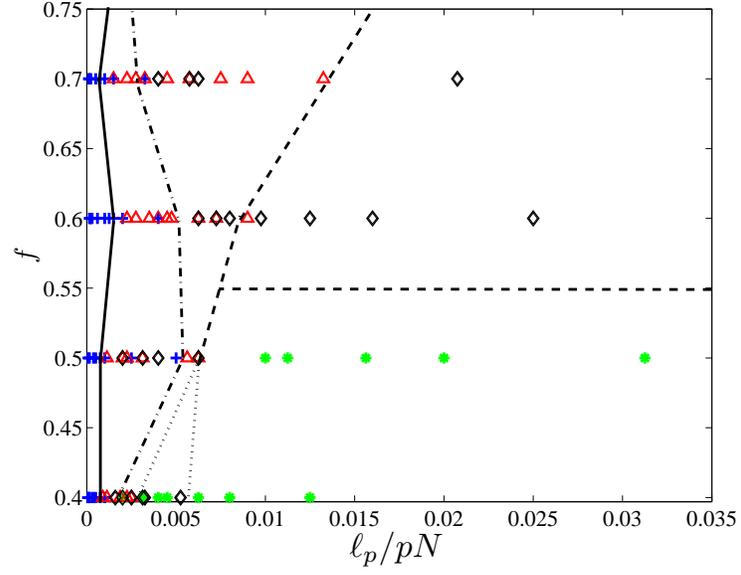}
\caption{Phase map for a $N = 400$ chain as a function of the hydrophobic fraction $f$, and the ratio of the persistence length to the periodicity length normalised by the total number of beads, $\ell_{p}/p N$. For small values of persistence length this corresponds to a flexible chain. The different phases globule ($+$), coil ($*$), folded ($\triangle$), and looped ($\diamond$) are shown. The lines indicate $50 \%$ stability lines for the different phases.}\label{fig:Phase-Map-400}
\end{figure}

Phase maps based on the equilibrium morphologies of the polymers for both cases, $N = 100$ and $N = 400$, are shown in Figs.~\ref{fig:Phase-Map-100} and \ref{fig:Phase-Map-400}, respectively. A complete listing of equilibrium structures for both $N = 100$ and $N = 400$, for all simulated values of $f$, $p$, and $\ell_{p}$, can be found in the Appendix (Tables~\ref{tab:Appendix-1} and \ref{tab:Appendix-2}). The phase maps indicate the occurence of a particular equilibrium structure as a function of the hydrophobic fraction, $f$, on the $Y-$axis and a dimensionless parameter, $\ell_{p}/(pN)$, on the $X-$axis. The dimensionless parameter, $\ell_{p}/(pN)$, reflects the ease of association of hydrophobic segments that are separated from each other along the chain contour by a characteristic distance.

In both cases, we find that a narrow globule region exists for very small values of $\ell_{p}/(pN)$ that spans the entire range of values of $f$. An increase in $\ell_{p}/(pN)$ results in a less compact structure with a larger equilibrium size. Thus, for $N = 100$, larger $\ell_{p}/(pN)$ values result in equiilbrium structures that are either coils or extended coils, and span most of the phase map. However, for certain intermediate values of $\ell_{p}/(pN)$ and large $f$, there is a narrow region where elongated folds are the equilibrium structures. In the case of $N = 100$, the ratio $\ell_{p}/L$, which defines the degree of flexibility of the polymer backbone, varies between 0.006 - 0.25. As a result, for large $\ell_{p}$ (where $\ell_{p} \sim L$), the polymer is semiflexible and
does not easily bend to form elongated folds or loops. This explains the complete absence of loops in Fig.~\ref{fig:Phase-Map-100}, and the existence of only a very narrow region for elongated folds.

For $N = 400$, $\ell_{p}/L$ varies between 0.001 and 0.0625 ($\ell_{p} << L$), which makes the polymer flexible over the length, $L = 400$. Regardless, an increase in $\ell_{p}$ leads to a decrease in chain flexibility. The resulting phase map (Fig.~\ref{fig:Phase-Map-400}), now shows distinct regions corresponding to all four structures, namely, globule, coil, elongated folds, and loops, respectively. These structures are distributed across the phase map in the following manner. Core-shell globules are observed for very small values of $\ell_{p}/pN$ along the left hand-side of the plot. The coil region is located in the lower right hand-side of the phase map corresponding to large values of $\ell_{p}/pN$ and small values of $f$. In contrast, loop structures are found predominantly in the upper right hand-side of the phase map, corresponding to large values of both $\ell_{p}/pN$ and $f$. The loop region also extends into a narrow band that lies between the globule and coil regions for small $f$ in the lower part of the phase map. Most of elongated fold structures exist in a region between the globule and loop regions for large $f$. Similar to the behavior of the loop region, the elongated fold region also extends to low values of $f$, and appears as a narrow band between globules and loops. The occurrence of elongated folds for $N = 400$ at lower values of $f$ ($f = 0.4, 0.5$) is in contrast with the phase map for $N = 100$, which is marked by an absence of elongated folded structures at $f = 0.4$ and $0.5$. This is because for $N = 400$, the polymer is less rigid, which makes it easier for hydrophobic segments to hybridize over longer lengths of the chain contour, and form folds even at low $f$. For $N = 100$, the chain is more rigid and can form folded structures only when a significant hydrophobic attraction energy at large $f$ is able to overcome the bending energy cost. Hence, we do not observe folded structures at low values of $f$.

Though different equilibrium configurations of polymers has been reported in this study, an order parameter that distinguishes the different phases have not been identified. The lines in Fig.~\ref{fig:Phase-Map-100} and \ref{fig:Phase-Map-400} demarcating different regions of parameter space where each phase is stable are guides to the eye drawn through the $50 \%$ relative stability for each of the phases. Thus the horizontal dashed line in Fig.~\ref{fig:Phase-Map-400} parallel to the $\ell_{p}/p N$ axis at a value of $f \approx 0.55$ demarcates the coil (indicated by $*$) and looped (indicated by $\diamond$) phases. A thorough analysis of the phase diagram indicating the lines of metastability and spinodal will be carried out in a future study.

\subsection{\label{sec:Forceext}Force vs.\ extension behavior}

\begin{figure}[ht]
\includegraphics[width=10cm]{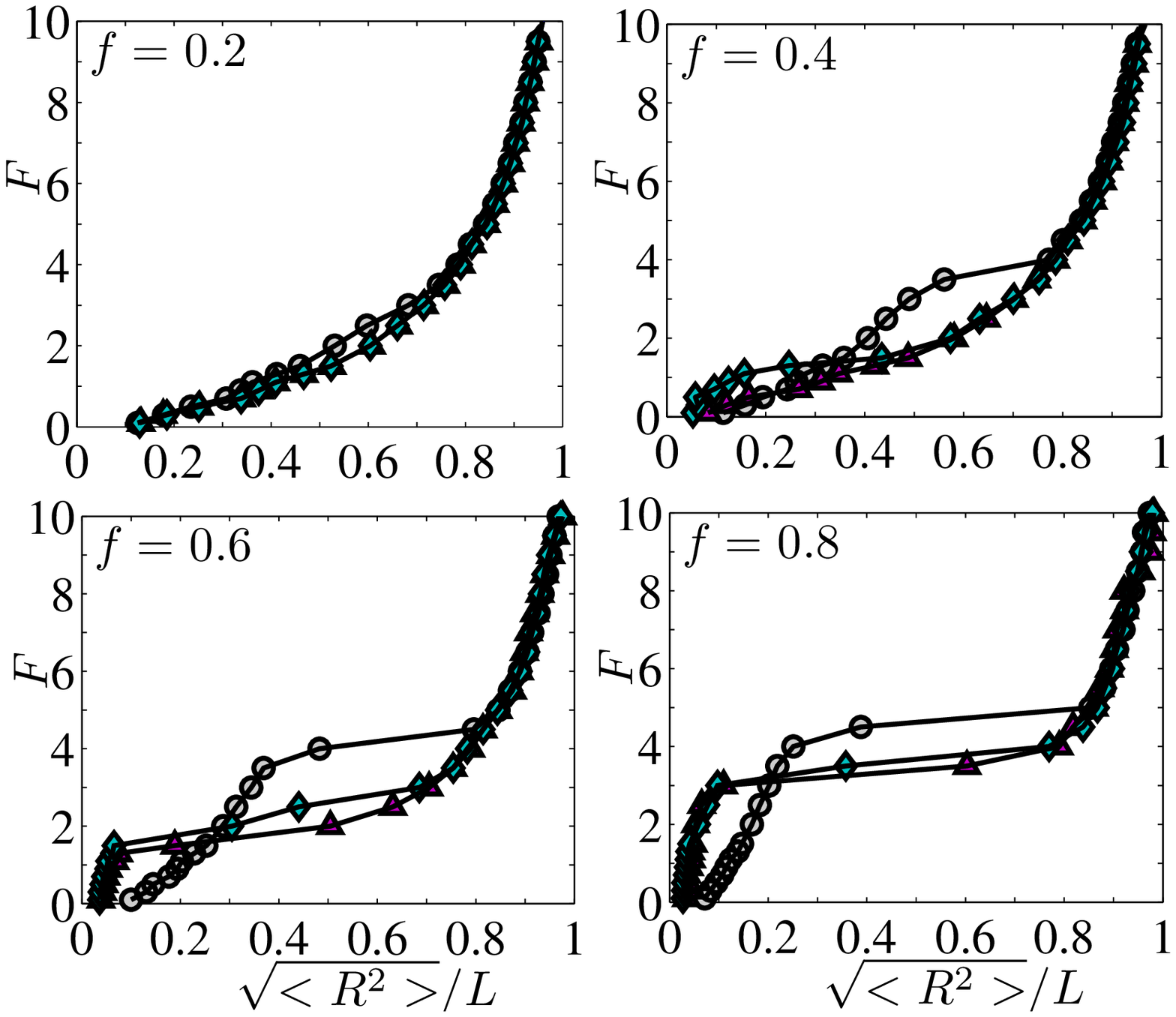}
\caption{Figure showing force, $F$, vs.\ extension, $\langle R^{2} \rangle^{1/2}/L$, curves corresponding to $N = 100$ copolymer chains for different values of the hydrophobic fraction, $f$, and different disorder realizations along the chain backbone, block copolymer (filled circles $\circ$), periodic (triangles $\triangle$), and random distribution (diamonds $\lozenge$).}
\label{fig:FvsExt}
\end{figure}

\begin{figure}[ht]
\includegraphics[width=10cm]{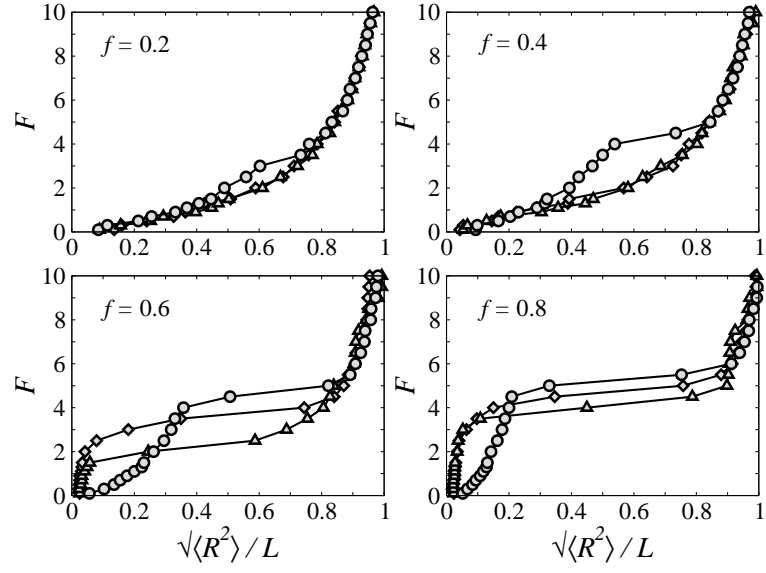}
\caption{Figure showing force, $F$, vs.\ extension, $\langle R^{2} \rangle^{1/2}/L$, curves corresponding to $N = 200$ copolymer chains for different values of the hydrophobic fraction $f$ and different disorder realizations along the chain backbone, block copolymer (filled circles $\circ$), periodic (triangles $\triangle$), and random distribution (diamonds $\lozenge$).}
\label{fig:FvsExt200}
\end{figure}

We now compare the force-extension response of copolymers with different distributions of hydrophobic groups. Figure~\ref{fig:FvsExt} shows force, $F$, vs. extension, $\left< R^{2}\right>^{1/2}/L$, for all three distributions at four different hydrophobic fractions, $f = 0.2$, $0.4$, $0.6$ and $0.8$ for $N = 100$. Representative force-extension curves for $N = 200$ for the same values of $f$ are presented in Fig.~\ref{fig:FvsExt200}. We would like to note two important aspects of the force-extension curves in Figs.~\ref{fig:FvsExt} and \ref{fig:FvsExt200}. First, the force-extension curves correspond to completely flexibile polymers with zero bending energy and $\ell_{p} = 0.6$. Second, the force-extension curves for the periodic distributions represent the smallest periodicity length, $p$, for a given value of $f$. The effects of both $\ell_{p}$ and $p$ on the force-extension behaviour are discussed in the later part of this section.

For small hydrophobic fraction, $f = 0.2$ (Fig.~\ref{fig:FvsExt}(a)), the chain extension increases monotonically with $F$ for all three distributions, with little variation between them.  For higher hydrophobic fractions $\left(f = 0.4, 0.6, 0.8\right)$, a first order globule to coil transition is observed at a critical force, $F_{c}$.
In the case of periodic and random distributions, the hydrophobic groups, which are distributed along the entire chain contour, hybridize with each other to form a compact globule. Therefore, when the magnitude of the applied force is small, the globule is only slightly distorted from its equilibrium conformation. Consequently, the extension remains nearly constant for small $F$. In contrast, the block copolymer undergoes a relatively larger extension for smaller values of $F$ ($F < F_{c}$) for $f = 0.4, 0.6, 0.8$. This is because in poor solvents, the block copolymer consists of a hydrophobic globule at the end of a flexible, hydrophilic coil that is easy to stretch under the application of a tensile force. Once the long wavelength fluctuations along the chain backbone have been smoothed out, the end-to-end distance of the chain remains constant upon an increase in the force $F$. However, beyond the critical force $F_{c}$, the chain unravels undergoing a jump discontinuity in size. On further increase in $F$ above $F_{c}$, the chain stretches out to full extension. As is expected, the large force behavior is identical for different disorder realizations as shown in Fig.~\ref{fig:FvsExt}.

We also note the mechanical response between periodic and random distributions for higher hydrophobic fractions $f = 0.6$, and $f = 0.8$ past the first order globule-coil phase transition is more pronounced as the chain length is increased from $N=100$ to $N=200$ (see Fig.~\ref{fig:FvsExt} and Fig.~\ref{fig:FvsExt200} panels(c) and (d)). Based on our simulations we conclude that the force-extension curve for the random distribution lies above that of the periodic ones for the parameters quoted here.

Comparing the force-extension curves for $N = 100$ and $200$ in Figs.~\ref{fig:FvsExt} and \ref{fig:FvsExt200}, respectively, we find that both are qualitatively the same. In other words, an increase in the chain length does not alter the force-extension response. In fact, we see the same response for the larger chain lengths, $N = 300$ and $400$, that we have simualated (results not shown here). Figures ~\ref{fig:FvsExt} and \ref{fig:FvsExt200} differ only in the values of $F_{c}$, which is slightly higher for $N = 200$ and also the response past the transition threshold as alluded to in the earlier paragraph. This can be expected because the overall binding energy of the globule would increase with increase in the number of hydrophobic groups.

\begin{figure}[ht]
\includegraphics[width=10cm]{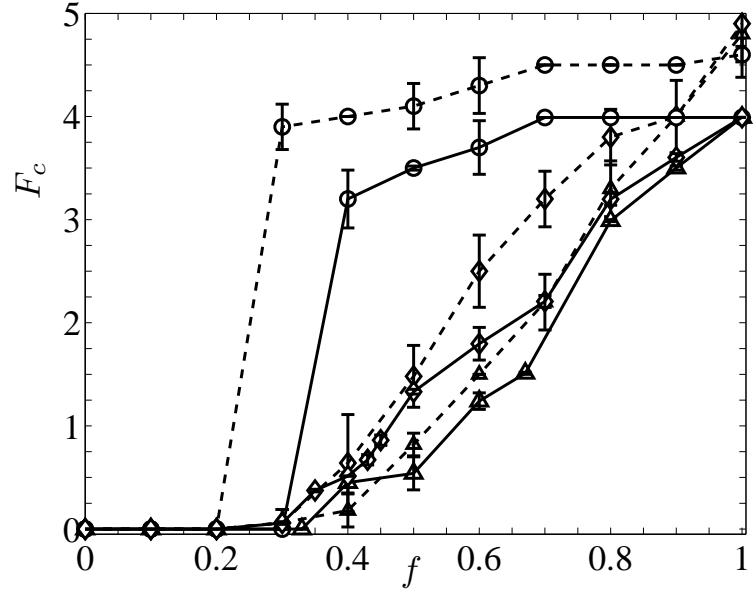}
\caption{Figure shows the variation of critical force $F_{c}$ as a function of the fraction of hydrophobic groups $f$ for different disorder realizations (a) block copolymer (circles $\circ$), (b) periodic distribution (triangles $\triangle$), and random distribution (diamonds $\lozenge$), and different chain lengths $N=100$ (solid line), and $N=200$ (dashed line). The average is over $5$ different realizations for the random system.}
\label{fig:Fcvsf}
\end{figure}

Figure~\ref{fig:Fcvsf} shows the variation of $F_{c}$ as a function of $f$ for $N=100, 200$ for all three distributions of hydrophobic groups. For $N=100$, and $f < 0.3$, the critical force, $F_{c}$, required to uncoil the polymer chain is zero for all three distributions. This is because for small $f$, the total number of hydrophobic contacts are few. Thus, the polymer remains in a coil-like state, and starts stretching continuously even for small values of $F$. In case of the block copolymer, $F_{c}$ increases discontinuously to a large value of approximately $3.2$ (from a zero value) for $f \approx 0.35$, and with increasing $f$, $F_{c}$ quickly saturates to a value of $4.0$. In contrast, for the periodic and random distributions, $F_{c}$ appears to increase linearly with $f$ for $f \gtrsim 0.35$. For $N=200$ chains and a block copolymer distribution of hydrophobic beads, the critical force is nonzero beyond a fraction of $\approx 0.2$. A finite size scaling analysis (not shown here) in which the critical hydrophobic fraction is plotted against $1/L$ yields a value of $f \approx 0.2$ in the thermodynamic limit $L \rightarrow \infty$. In contrast, the critical fraction remains $\approx 0.35$ for periodic and random distributions for both $N=100$ and $N=200$ chains.

\begin{figure}[ht]
\includegraphics[width=10cm]{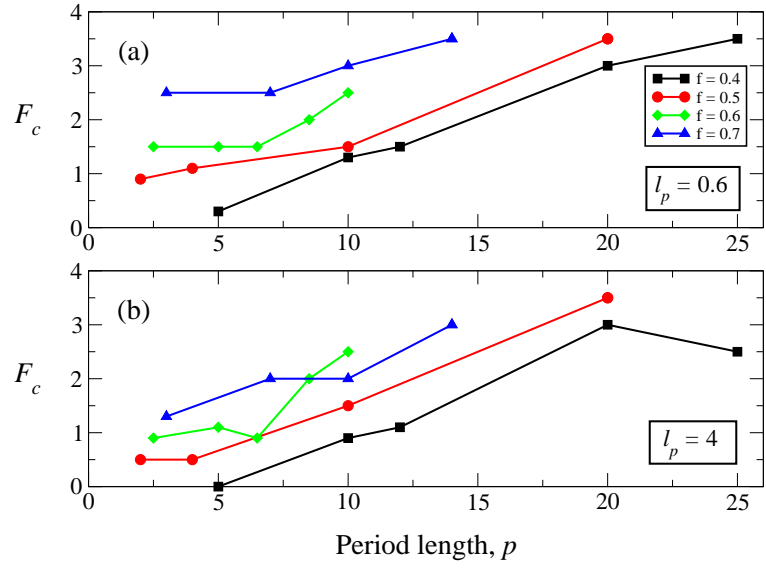}
\caption{Variation of the critical force, $F_{c}$, as a function of periodicity length, $p$, for different hydrophobic fractions, $f$, and small values of $\ell_{p}$ for $N = 400$ copolymers; (a) $\ell_{p} = 0.6$, and (b) $\ell_{p} = 4$.}
\label{fig:Fcvsp-Smallp}
\end{figure}

As we have seen in Figs.~\ref{fig:Rep-Phase-400} $-$ \ref{fig:Phase-Map-400}, both $\ell_{p}$ and $p$ have a profound impact on the equilibrium description of the copolymer. Hence, we can expect that the force-extension response will also be influenced by a finite bending energy, and the disorder correlation length along the polymer backbone. Figure~\ref{fig:Fcvsp-Smallp} describes the variation of the critical transition force, $F_{c}$, with the periodicity length, $p$, for different values of $f$, for a chain length of $N = 400$. These $F_{c}$ values correspond to two of the smaller persistence lengths, $\ell_{p} = 0.6$ (Fig.~\ref{fig:Fcvsp-Smallp}(a)) and $\ell_{p} = 4$ (Fig.~\ref{fig:Fcvsp-Smallp}(b)) considered in our simulations. The $F_{c}$ values for Fig.~\ref{fig:Fcvsp-Smallp} have been estimated from individual force-extension curves corresponding to a specific combination of values of $f$, $p$, and $\ell_{p}$. We know from Fig.~\ref{fig:Rep-Phase-400} that the equilibrium phase at small $\ell_{p}$ is a globule (except for $f = 0.4$, $p = 5$, $\ell_{p} = 4$ where it is a coil), and the force-extension curves show a single globule-coil transition at some force, $F = F_{c}$. For both $\ell_{p} = 0.6$ (zero bending energy) and $\ell_{p} = 4$, $F_{c}$ increases almost linearly with $p$ for a given value of $f$.  If $\ell_{p}$ is small, then the polymer chain bends easily bringing different hydrophobic patches along the contour closer. For a constant $f$, a larger value of $p$ implies that longer stretches of hydrophobic beads are found along the chain backbone. A longer hydrophobic stretch would provide a larger ``contact area" for another hydrophobic stretch that binds to it, thus increasing both contact probability and the binding energy. This increase would manifest itself as an increase in $F_{c}$ for a larger $p$. In addition, as one would expect from intuition, the critical force also increases with $f$ for a fixed value of $p$.

\begin{figure}[ht]
\includegraphics[width=10cm]{Figure-11a.eps}
\newline
\includegraphics[width=10cm]{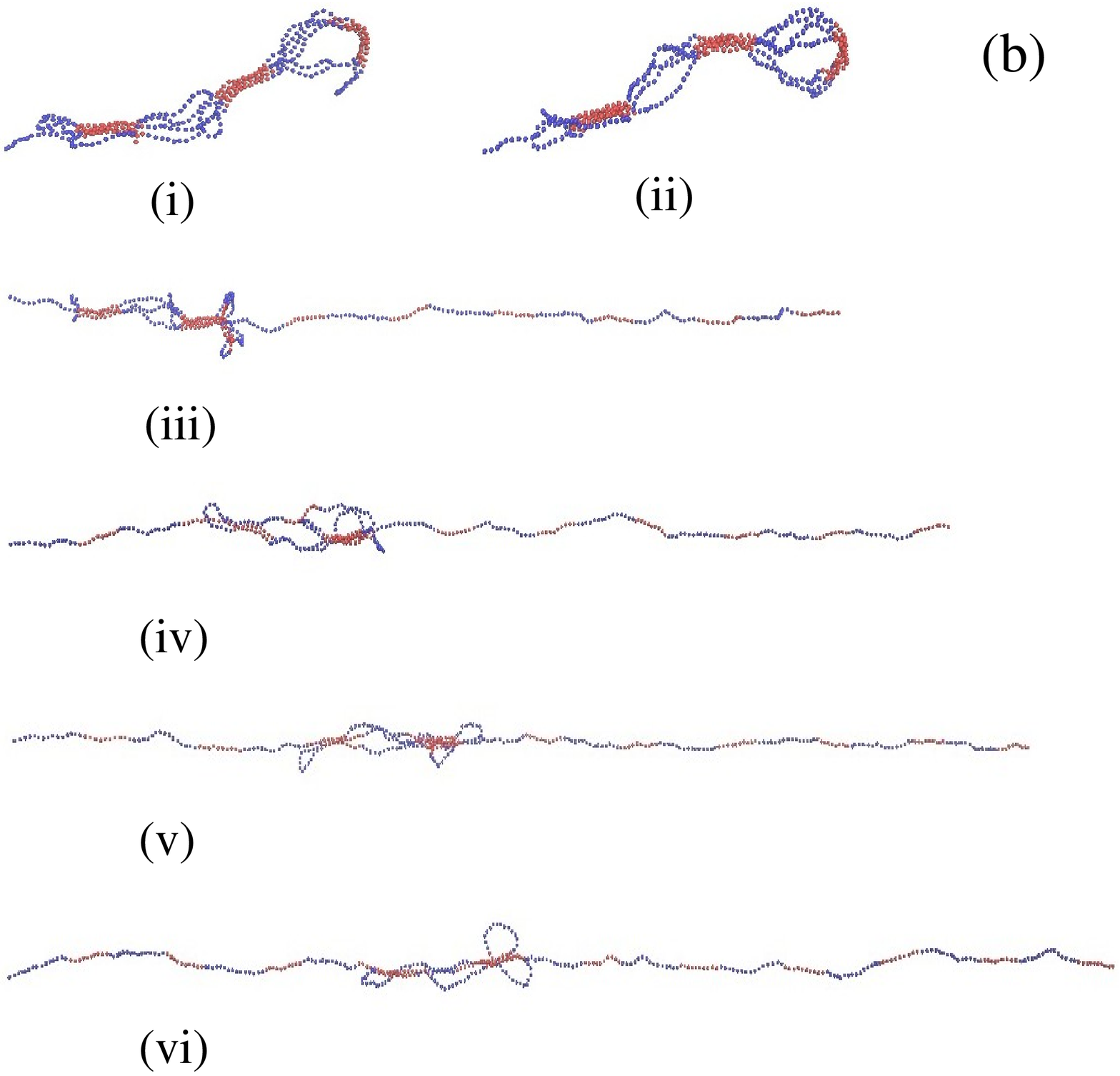}
\caption{Figure shows, (a) the force-extension curve with multiple plateaus for an elongated fold configuration for $N = 400$, $f = 0.4$, $p = 25$, and $\ell_{p} = 9$, and (b) configurations of the elongated fold at various stages of its unfolding corresponding to the multiple plateaus in the force-extension curve.}
\label{fig:Fvsx-Fold}
\end{figure}

\begin{figure}[ht]
\includegraphics[width=10cm]{Figure-12a.eps}
\includegraphics[width=8cm]{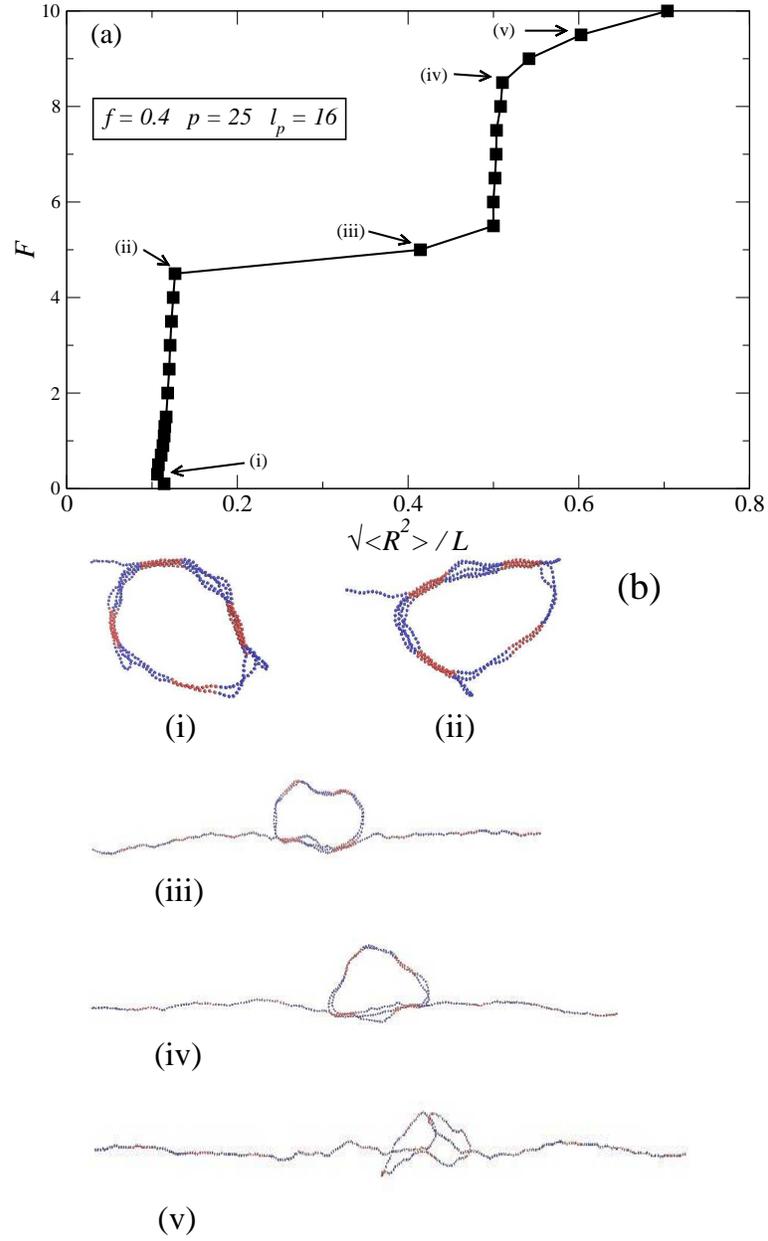}
\caption{Figure shows, (a) the force extension curve with multiple plateaus for a loop configuration for $N = 400$, $f = 0.4$, $p = 25$, and $\ell_{p} = 16$, and (b) configurations of the loop at various stages of its unfolding corresponding to the multiple plateaus in the force-extension curve. The transition between snapshots (iv) and (v) clearly shows the de-binding of the final two strands that make the loop structure.}
\label{fig:Fvsx-Loop}
\end{figure}

A simpler characterization of the force-extension curves becomes more difficult for larger persistence lengths. Equilibrium structures at large $\ell_{p}$ values are either elongated folds or loops, organized as several strands folded on top of each other and held together by stretches of hydrophobic beads. In contrast with force-extension curves for small $\ell_{p}$ globular structures which show a single plateau at $F = F_{c}$ corresponding to a globule-coil transition, a typical force-extension curve for large $\ell_{p}$ values shows several plateaus corresponding to succesive unbinding of strands from the folded equilibrium structure. As a result it becomes difficult to define a single value of $F_{c}$. Such multiple transitions are observed for elongated folds and loops in our simulations, and are shown in Figs.~\ref{fig:Fvsx-Fold} and \ref{fig:Fvsx-Loop} for an elongated fold and loop, respectively.

The force-extension curve in Fig~\ref{fig:Fvsx-Fold}(a) represents the unfolding and stretching of an elongated fold structure ($N = 400$, $f = 0.4$, $p = 25$, $\ell_{p} = 9$). At least four transitions (force plateaus) can be spotted at $F = 2.0, 5.5, 6.5, 8.5$ and $9.5$ for the range of $F$ values considered in our simulations, and these events are marked from (i) - (vi). The equivalent configuration snapshots corresponding to the events in Fig~\ref{fig:Fvsx-Fold}(a) are shown in Fig~\ref{fig:Fvsx-Fold}(b). We can see a large change in both size and structure between events (ii) and (iii), which is also evident from the corresponding size change in the force-extension curve. Although, successive transitions are not accompanied by large changes in chain extension, snapshots (iii) - (vi) do show a step-wise unfolding of the elongated fold structures in (i) and (ii).

Figure~\ref{fig:Fvsx-Loop}(a) shows that the force-extension response of an equilibrium loop structure ($N = 400$, $f = 0.4$, $p = 25$, $\ell_{p} = 16$) is similar to that of an elongated fold structure in Fig~\ref{fig:Fvsx-Fold}(a), with multiple transitions at $F = 4.5, 5.5,$ $and 8.5$. We would like to note here that only a change in $\ell_{p}$ between Figs.~\ref{fig:Fvsx-Fold} and \ref{fig:Fvsx-Loop} has led to a transition from an elongated fold to a loop structure. From the configuration snapshots in Figure~\ref{fig:Fvsx-Loop}(b), we see a stepwise peeling of the loop structure. Between (ii) and (iii), we see a large change in extension with opening up of one of the hybridized strands into a straight coil. The subsequent change from stage (iii) - (iv) is accompanied by further stretching of the coil, while the loop remains intact. However, between stages (iv) and (v), we can clearly see the dissociation of the final two strands that form the loop structure.

\section{\label{sec:summary}Summary}

We have utilized Langevin dynamics simulations to investigate the equilibrium morphologies and force-extension behavior of a single copolymer composed of hydrophobic and hydrophilic groups arranged in periodic, random and block distributions along its backbone. Our simulations demonstrate that in addition to the fraction of hydrophobic groups on the chain, a competition between various length scales, namely, the persistence length, the periodicity length (\textit{i.e.} disorder correlation length) in comparison to the total chain length play an important role in dictating the equilibrium morphologies. For small persistence lengths and long chains, we find a core-shell equilibrium morphology with the hydrophilic beads on the outside. In case the polymer chain is small, similar core-shell morphologies, with some hydrophobic beads exposed are seen. This is due to the semi-flexibility of the chain that makes it energetically unfavorable to bend the chain over length scales smaller compared to its persistence length. The semi-flexibility manifests itself at higher disorder fractions giving rise to looped conformations with hybridised hydrophobic/hydrophilic beads in the low persistence length, long chain limit, and extended coil conformations in the short chain limit. The plethora of phases arising in this system has been summarized in a phase map.

We further report that the mechanical response of a heteropolymer undergoing a force-induced globule-coil transition is dependent on the distribution of hydrophobic groups along its backbone. For flexible chains all three disorder realizations have globular equilibrium morphologies at large hydrophobic fractions, and show a first order transition in their sizes at a critical force, $F_{c}$. However, for a given fraction, the critical force is observed to be higher for the block copolymer in comparison to the periodic and random distributions. In addition, our simulations show that the force-extension responses of the periodic and random distributions are nearly identical for short chains ($N = 100$). For longer chains ($N = 200 - 400$) a difference in force-extension behaviors past the transition point is seen. We further plotted the critical force, $F_{c}$, as a function of the hydrophobic fraction, $f$, for different disorder realisations. Since the polymers behave like coils for small hydrophobic fractions, $F_{c} = 0$ for all three distributions. An important result is that $F_{c}$ is non-zero for $f \approx 0.35$ for the random and periodic distributions, which is different from $f \approx 0.2$ for block copolymers. On further increase in the hydrophobic fraction, the critical force shows a first order transition for the block copolymer. However, it increases continuously for the random and periodic distributions.

For semi-flexible chains with small persistence lengths ($l_{p} = 0.6, 4$), we observe a core-shell to coil transition in the force-extension behavior corresponding to a single unfolding event. The critical force, $F_{c}$, which characterizes this unfolding event is found to increase almost linearly with the disorder correlation length, $p$. However, as the persistence length increases more complex folded structures (elongated folds and loops) are observed at equilibrium. As a consequence, the force-extension behavior shows a cascade of transitions corresponding to successive unfolding of the hybridized domains.

We now place our results in context of existing theories on the subject. The collapse transition of disordered polymers has been the subject of several studies~\cite{Grosberg:85,Obukhov:86,Garel:94}. Statistical techniques rooted in replica methods has been used to explore globular conformations~\cite{Pande:00} (and references therein). Further optimal distributions of hydrophobic and hydrophilic groups along the primary sequence leading to folded conformations have been investigated in context of rational design of proteins~\cite{Chertovich:00,Kriksin:02,Zheligovskaya:99,Khokhlov:99,vandanOever:99,Berezkin:03,Khokhlov:04,Chertovich:04,Lau:89,Yue:92}. and statistical mechanical theories of random systems~\cite{Grosberg:85,Garel:94} have been employed to explore such conformational states of A-B/H-P heteropolymers.

However, mechanical behavior of disordered polymers have been less investigated~\cite{Bensimon:98}. Our results illustrate the difference in mechanical properties of such heteropolymers having similar bulk structural properties but differing in the local arrangement of disorder undergoing a force induced globule-coil transition. Transfer matrix calculations based on the helix-coil model for annealed disorder~\cite{Chakrabarti:05,Chakrabarti:06} can be adapted to the case of quenched disorder. However, this is omitted from the present paper for brevity and the fact that it does not yield any new insights over those obtained from simulations.

This work opens up several avenues of theoretical research. In particular, an order parameter based characterisation of the menagerie of observed phases, the transition between these phases and further the dynamics of tension induced unravelling of hierarchically folded structures. So far such tension induced dynamic transitions of globules~\cite{Katz:09,Sing:11,Einert:11} have been reported only for homopolymers, and would need to be adapted for the specific case studied here. We believe that a more generic model that has been used to study such dynamical transitions in context of polyelectrolytes~\cite{Panwar:09} will prove handy for the purpose.

We hope that our work will spark interest amongst polymer chemists interested in rational design of polymers that aim to identify sequences that lead to specific folded structures. Further, we believe this work would also be useful to single molecule biophysicists~\cite{Sing:12} interested in understanding protein conformational transitions and allostery.

\begin{appendices}
\section{Appendix}

Tables~\ref{tab:Appendix-1} and \ref{tab:Appendix-2} contain the complete list of equilibrium structures for both $N = 100$ and $N = 400$, for all simulated values of $f$, $p$, and $\ell_{p}$.

\end{appendices}

\newpage

\newpage
\section{Figure captions}
\noindent{\bf{Fig. 1:}} Schematic representation showing hydrophobic groups (black circles) arranged along the chain backbone in (a) random, (b) periodic, and (c) block copolymer arrangements respectively. The fraction of hydrophobic beads is $f = 0.5$.

\noindent{\bf{Fig. 2:}} shows the average radius of gyration, $\langle R_g \rangle$, for a $N = 100$ chain over different disorder realizations as a function of the hydrophobic groups for (a) block copolymer (filled circles $\circ$), (b) periodic distribution (filled triangles $\triangle$), and random distribution (open diamonds $\lozenge$). The average is over $5$ different realizations for the random system.

\noindent{\bf{Fig. 3:}} Equilibrium morphologies of a $N = 400$ polymer chain as a function of different hydrophobic fractions, $f$, periodicity lengths, $p$, and persistence lengths, $\ell_{p}$. The hydrophobic and hydrophilic beads are indicated in red and blue colors, respectively. The chain rigidity/persistence length increases along each row while the hydrophobic fraction increases along each column. For flexible chains a core-shell like morphology is seen for all values of $f$, while for semi-flexible chains with large periodicity lengths, looped configurations are observed.

\noindent{\bf{Fig. 4:}} Representative snapshots of dressed polymer architectures for $N = 100$ chain as a function of different hydrophobic fractions $f$, periodicity lengths $p$, and persistence length $\ell_{p}$. While core-shell morphologies similar to the $N = 400$ case is observed, looped configurations are rarely seen even for large persistence and periodicity lengths.

\noindent{\bf{Fig. 5:}} Phase map for a $N = 100$ chain. The two key parameters that dictate the stability of different configurations is the hydrophobic fraction $f$, and the ratio of the persistence length to the periodicity length normalised by the total number of beads, $\ell_{p}/p N$. The different phases are indicated by globule ($+$), coil ($*$), folded ($\triangle$), and extended coils ($\circ$). The lines indicate $50 \%$ stability lines for the different phases.

\noindent{\bf{Fig. 6:}} Phase map for a $N = 400$ chain as a function of the hydrophobic fraction $f$, and the ratio of the persistence length to the periodicity length normalised by the total number of beads, $\ell_{p}/p N$. For small values of persistence length this corresponds to a flexible chain. The different phases globule ($+$), coil ($*$), folded ($\triangle$), and looped ($\diamond$) are shown. The lines indicate $50 \%$ stability lines for the different phases.

\noindent{\bf{Fig. 7:}} Figure showing force, $F$, vs.\ extension, $\langle R^{2} \rangle^{1/2}/L$, curves corresponding to $N = 100$ copolymer chains for different values of the hydrophobic fraction, $f$, and different disorder realizations along the chain backbone, block copolymer (filled circles $\circ$), periodic (triangles $\triangle$), and random distribution (diamonds $\lozenge$).

\noindent{\bf{Fig. 8:}} Figure showing force, $F$, vs.\ extension, $\langle R^{2} \rangle^{1/2}/L$, curves corresponding to $N = 200$ copolymer chains for different values of the hydrophobic fraction $f$ and different disorder realizations along the chain backbone, block copolymer (filled circles $\circ$), periodic (triangles $\triangle$), and random distribution (diamonds $\lozenge$).

\noindent{\bf{Fig. 9:}} Figure shows the variation of critical force $F_{c}$ as a function of the fraction of hydrophobic groups $f$ for different disorder realizations (a) block copolymer (circles $\circ$), (b) periodic distribution (triangles $\triangle$), and random distribution (diamonds $\lozenge$), and different chain lengths $N=100$ (solid line), and $N=200$ (dashed line). The average is over $5$ different realizations for the random system.

\noindent{\bf{Fig. 10:}} Variation of the critical force, $F_{c}$, as a function of periodicity length, $p$, for different hydrophobic fractions, $f$, and small values of $\ell_{p}$ for $N = 400$ copolymers; (a) $\ell_{p} = 0.6$, and (b) $\ell_{p} = 4$.

\noindent{\bf{Fig. 11:}} Figure shows, (a) the force-extension curve with multiple plateaus for an elongated fold configuration for $N = 400$, $f = 0.4$, $p = 25$, and $\ell_{p} = 9$, and (b) configurations of the elongated fold at various stages of its unfolding corresponding to the multiple plateaus in the force-extension curve.

\noindent{\bf{Fig. 12:}} Figure shows, (a) the force extension curve with multiple plateaus for a loop configuration for $N = 400$, $f = 0.4$, $p = 25$, and $\ell_{p} = 16$, and (b) configurations of the loop at various stages of its unfolding corresponding to the multiple plateaus in the force-extension curve. The transition between snapshots (iv) and (v) clearly shows the de-binding of the final two strands that make the loop structure.

\newpage
\section{Table Captions}

\noindent{\bf{Table A-1:}} Table showing different equilibrium structures obtained for $N = 100$ for all simulated values of $f$, $p$, and $\ell_{p}$.

\noindent{\bf{Table A-2:}} Table showing different equilibrium structures obtained for $N = 400$ for all simulated values of $f$, $p$, and $\ell_{p}$.

\newpage
\newpage
\newpage
\newpage
\newpage
\newpage
\newpage
\newpage
\newpage
\newpage
\newpage
\newpage
\newpage
\begin{table}[ht]
\includegraphics[width=12cm]{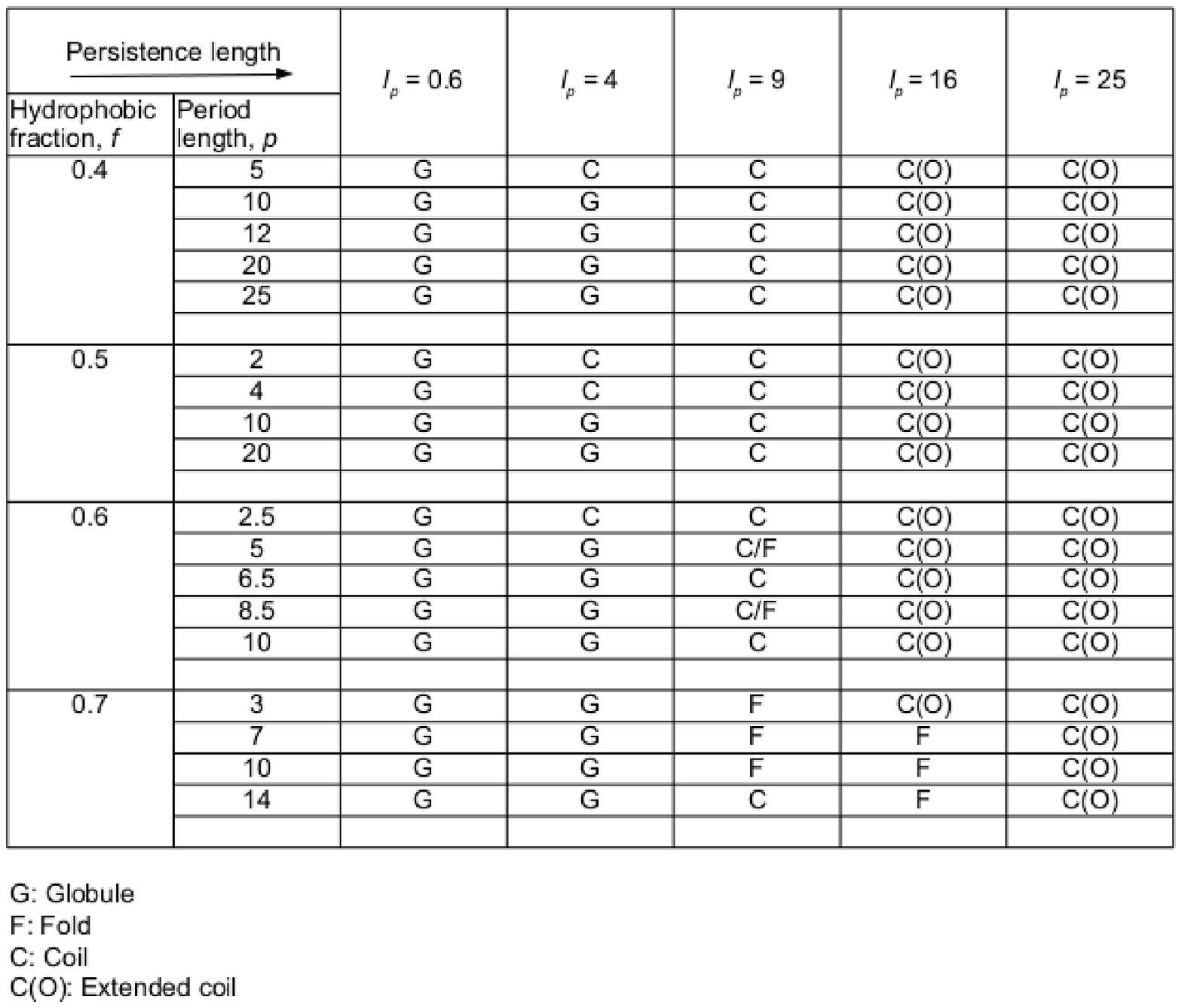}
\caption{Table showing different equilibrium structures obtained for $N = 100$ for all simulated values of $f$, $p$, and $\ell_{p}$.}
\label{tab:Appendix-1}
\end{table}
\newpage
\begin{table}[ht]
\includegraphics[width=12cm]{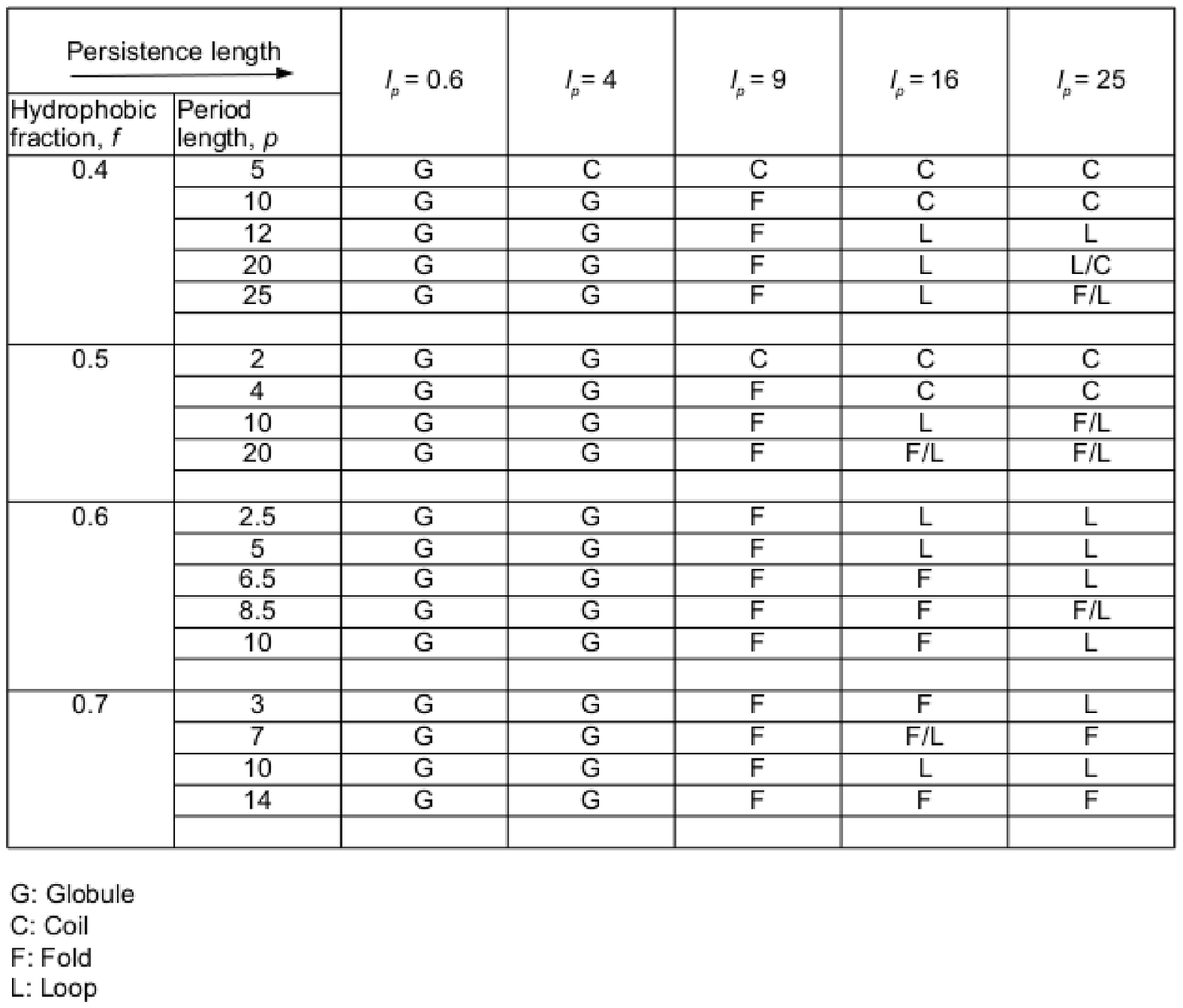}
\caption{Table showing different equilibrium structures obtained for $N = 400$ for all simulated values of $f$, $p$, and $\ell_{p}$.}
\label{tab:Appendix-2}
\end{table}


\begin{thebibliography}{99}

\section{References}

\bibitem{Alberts:02} B. Alberts, A. Johnson, J. Lewis, M. Raff, K. Roberts, and P. Walter, \textit{Molecular Biology of the Cell}, Garland, New York (2002).

\bibitem{Bustamante:11} C. Bustamante, W. Cheng, and Y. X. Mejia, Cell, {\bf 144}, 480 (2011).

\bibitem{Poland:70} D. Poland, and D. Scheraga, \textit{Theory of Helix-Coil transitions in biopolymers: statistical mechanical theory of order-disorder transition in biological macromolecules}, Academic Press (1970).

\bibitem{Fisher:64} M. E. Fisher, Am. J. Phys. {\bf 32}, 343 (1964).

\bibitem{Marko:95} J. Marko, and E. Siggia, Macromol. {\bf 28}, 8759 (1995).

\bibitem{Kierfeld:04} J. Kierfeld, O. Niamploy, V. Sa-Yakanit, and R. Lipowsky, Euro Phys. J. E., {\bf 14}, 17 (2004).

\bibitem{Chakrabarti:05} B. Chakrabarti, and A. J. Levine, Phys. Rev. E {\bf 71}, 031905 (2005).

\bibitem{Chakrabarti:06} B. Chakrabarti, and A. J. Levine, Phys. Rev. E {\bf 74}, 031903 (2006).

\bibitem{Chertovich:00} A. V. Chertovich, V. A. Ivanov, A. A. Lazutin, and A. R. Khokhlov, Macromol. Symp. {\bf 160}, 41 (2000).

\bibitem{Kriksin:02} Y. A. Kriksin, P. G. Khalatur, and A. R. Khokhlov, Macromol. Theory Simul. {\bf 11}, 213 (2002).

\bibitem{Zheligovskaya:99} E. A. Zheligovskaya, P. G. Khalatur, and A. R. Khokhlov, {\bf 59}, 3071 (1999).

\bibitem{Khokhlov:99} A. R. Khokhlov, and P. G. Khalatur, Phys. Rev. Lett. {\bf 82}, 3456 (1999).

\bibitem{vandanOever:99} J. M. P. van den Oever, \textit{et al.} {\bf 65}, 041708 (1999).

\bibitem{Berezkin:03} A. V. Berezkin, P. G. Khalatur, and A. R. Khokhlov, J. Chem. Phys. {\bf 118}, 8049 (2003).

\bibitem{Khokhlov:04} A. R. Khokhlov, and P. G. Khalatur, Curr. Opin. Solid State Mater. Sci. {\bf 8}, 3 (2004).

\bibitem{Chertovich:04} A. V. Chertovich, E. N. Govorun, V. A. Ivanov, P. G. Khalatur, and A. R. Khokhlov, Eur. Phys. J. E {13}, 15 (2004).

\bibitem{Pande:00} V. S. Pande, A. Yu Grosberg, and T. Tanaka, Rev. Mod. Phys., {\bf 72}, 259 (2000).

\bibitem{Lau:89} K. F. Lau, and K. A. Dill, Macromol. {\bf 22} 3986 (1989).

\bibitem{Yue:92} K. Yue, and K. A. Dill, Proc. Natl. Acad. Sci. {\bf 89}, 4163 (1992).

\bibitem{Dasmahapatra:06} A. K. Dasmahapatra, G. Kumaraswami, and H. Nanavati, Macromol., {\bf 39}, 9621 (2006).

\bibitem{Dasmahapatra:07} A. K. Dasmahapatra, H. Nanavati, and G. Kumaraswami, J. Chem. Phys. {\bf 127}, 234901 (2007).

\bibitem{Shakhnovich:02} P. L. Geissler and E. I. Shakhnovich, Phys. Rev. E, {\bf 65}, 056110 (2002).

\bibitem{Shakhnovich2:02} P. L. Geissler and E. I. Shakhnovich, Macromolecules {\bf 35}, 4429, (2002).

\bibitem{vandenBroek:10} B. van den Broek, M. C. Noom, J. van Mameren, C. Battle, F. C. Mackintosh, and G. L. Wuite, Biophys. J. {\bf 98}, 1902 (2010).

\bibitem{Liu:02} S. Liu, and M. Muthukumar, J. Chem. Phys., {\bf 116}, 9975 (2002).

\bibitem{Plimpton:95} S. J. Plimpton, J. Comp. Phys., {\bf 117}, 1 (1995), \url{http://lammps.sandia.gov}.

\bibitem{Grosberg:85} A. Yu Grosberg, J. Stat. Phys., {\bf 38}, 149 (1985).

\bibitem{Obukhov:86} S. P. Obukhov, J. Stat. Phys., {\bf 19}, 3655 (1986).

\bibitem{Garel:94} T. Garel, L. Leibler, and H. J. Orland, J. Physique II, {\bf 4}, 2139 (1994).

\bibitem{Bensimon:98} D. Bensimon, D. Dohmi, and M. Mezard, Europhys. Lett., {\bf 42}, 97 (1998).

\bibitem{Einert:11} T. R. Einert, C.E. Sing, A. Alexander-Katz, R.R. Netz, Eur. Phys. J. E. {\bf 34}, 130 (2011).

\bibitem{Sing:11} C. E. Sing, T. R. Einert, R. R. Netz, and A. Alexander-Katz, Phys. Rev. E {\bf 83}, 040801(R) (2011).

\bibitem{Katz:09} A. Alexander-Katz, H. Wada, and R. R. Netz, Phys. Rev. Lett. {\bf 103}, 028102 (2009).

\bibitem{Panwar:09} A. S. Panwar, M. Kelly, B. Chakrabarti, and M. Muthukumar, \url{http://arxiv.org/abs/0905.4524}

\bibitem{Sing:12} C. E. Sing and A. Alexander-Katz, Macromol. {\bf 45}, 6704 (2012).

\end{thebibliography}
\end{document}